\documentclass[12pt]{article}

\usepackage{newtxtext,newtxmath}

\usepackage{graphicx}

\usepackage[letterpaper,margin=1in]{geometry}

\renewenvironment{abstract}
	{\quotation}
	{\endquotation}

\date{}

\makeatletter
\renewcommand{\fnum@figure}{\textbf{Figure \thefigure}}
\renewcommand{\fnum@table}{\textbf{Table \thetable}}
\makeatother

\usepackage{scicite}

\usepackage{url}

\def\scititle{
	Fully gapped superconductivity in magic-angle graphene Cooper-pair islands
}
\title{\bfseries \boldmath \scititle}

\author{
	Alexandra~Mestre-Tor\`a$^{1,2\ast}$,
	Marta~Perego$^{1,2}$,
	Clara~Galante~Agero$^{1,2}$,
	Takashi~Taniguchi$^{3}$,\and
	Kenji~Watanabe$^{4}$,
	Thomas~Ihn$^{1,2}$,
	Klaus~Ensslin$^{1,2}$
	Artem~O.~Denisov$^{1,2}$\and
	\small$^{1}$Laboratory for Solid State Physics, ETH Z\"urich, Z\"urich, Switzerland.\and
	\small$^{2}$Quantum Center, ETH Z\"urich, Z\"urich, Switzerland.\and
	\small$^{3}$Research Center for Materials Nanoarchitectonics, National Institute for Materials Science, Tsukuba, Japan.\and
	\small$^{4}$Research Center for Electronic and Optical Materials, National Institute for Materials Science, Tsukuba, Japan.\and
	\small$^\ast$Corresponding author. Email: amestre@phys.ethz.ch
}

\begin{document} 

\maketitle

\begin{abstract} \bfseries \boldmath

The mechanism behind superconductivity in magic-angle graphene systems remains an open question. Recent tunnelling and superfluid stiffness measurements have suggested a nodal superconducting gap, but these probes are typically applied to large-area devices, where twist-angle inhomogeneity and disorder can hinder resolving the intrinsic superconducting spectrum. 
Here we introduce quantum dots in magic-angle twisted bilayer graphene and access a regime where gate-tunable superconductivity competes with strong Coulomb repulsion upon adding an additional carrier to the island. 
The islands exhibit $2e$-periodic Coulomb blockade, revealing a well-defined quasiparticle gap in the island spectrum, inconsistent with the presence of nodes in the order parameter.
Within the same device, we also realize coupled superconducting islands and proximitized quantum dots, establishing a monolithic platform for hybrid dot devices. 
Our results point to a nodeless quasiparticle gap in moiré graphene and demonstrate the potential of mesoscopic hybrid structures to probe their correlated phases.

\end{abstract}

\noindent

\subsection*{\label{sec1:introduction} Introduction}

Magic-angle graphene enables the study of superconductivity in a strongly correlated flat-band system with a high degree of tunability~\cite{Cao2018, Cao2018_CI, Lu2019, Yankowitz2019}. 
Despite extensive experimental research, the microscopic origin, quasiparticle excitation spectrum and pairing symmetry of the superconducting state remain to be understood.
Early scanning tunneling microscopy~\cite{Oh2021, Kim2022}, together with more recent simultaneous transport and tunneling spectroscopy~\cite{Park2026} and superfluid stiffness measurements~\cite{Tanaka2025, Banerjee2025,Portoles2025}, reported signatures consistent with the presence of nodal superconductivity, where the gap in the
quasiparticle spectrum vanishes at isolated points or lines on the Fermi surface.
In contrast, new tunneling microscopy data resolved a fully opened superconducting gap around zero energy~\cite{Kim2026}.
This contrast may reflect the different length scales probed by the respective techniques. 
While tunnelling microscopy accesses local spectral features, transport measurements average over extended areas and are more susceptible to spatial inhomogeneity intrinsic to moiré systems~\cite{Uri2020, maccari2026}.

Here, we introduce gate-defined quantum dots in MATBG, enabling a mesoscopic transport probe that bridges the gap between local spectroscopy and transport techniques on extended systems.
By confining superconductivity to the few-hundred-nanometer scale, we access a regime where superconductivity competes with Coulomb repulsion.
In this limit, the superconducting parity effect becomes a transport probe for the low-energy quasiparticle states of the superconductor~\cite{lafarge1993,mineev1994, Higginbotham2015}: a fully opened gap protects Cooper-pair (charge $2e$) transport, whereas the presence of low-energy quasiparticle states poison the island and enable single-electron (charge $e$) transport.

We first show the formation of a superconducting island and observe $2e$-periodic Coulomb blockade, which evolves into $1e$-periodic transport when superconductivity is suppressed by a magnetic field or temperature.
This transport signature indicates a fully opened gap, as in conventional $s$-wave superconductors~\cite{Albrecht2016,Shen2018}.
Using the same device, we also realize the complementary regime of a normal quantum dot proximitized by superconducting MATBG leads.
In this regime, we measure an even-odd spectrum of Andreev bound states with zero-bias conductance peaks and a strong gate modulation of the supercurrent that can be understood in the context of a $0$-$\pi$ junction~\cite{vanDam2006}. 
Beyond the investigation of these quantum transport phenomena, we demonstrate a scalable extension from single to double quantum dots by increasing the number of finger gates.
This monolithic and reconfigurable architecture also overcomes the interface limitation of traditional superconducting–semiconducting systems~\cite{vanLoo2023, Pita-Vidal2023}.

\subsection*{\label{sec2:device} Device}
Figure~\ref{fig1}a shows a schematic of the device. The van der Waals heterostructure consists of MATBG encapsulated in hexagonal boron nitride (hBN) and a graphite flake that serves as a global back gate (BG). The stack is etched into a Hall-bar geometry with a channel width of 550 nm (Extended Data Fig.~\ref{figED1}). A thin layer of AlO$_{x}$ insulates the exposed edges of MATBG from the metallic top-gate layer, which consists of two finger gates (FGs) with a width and pitch of approximately ${150}~\textrm{nm}$. Instead of split gates, we use physical etching to define the channel, thereby avoiding potential current leakage beneath the gated regions due to percolation paths arising from twist-angle disorder~\cite{Uri2020}.
To reduce the impact of etch-induced disorder at the channel edges~\cite{jeong2025}, we minimize the width of the FGs so that the resulting elongated electrostatic potential profile ensures that the quantum dots are predominantly gate-defined.

We perform transport measurements in a dilution cryostat at a base temperature of 10 mK. Figure~\ref{fig1}b shows the four-probe longitudinal resistance $R$ of the bulk on a logarithmic scale as a function of back-gate voltage $V_{\textrm{BG}}$, moiré filling factor $\nu_\mathrm{BG}$ and temperature $T$. The MATBG is twisted at an angle of $\theta \approx 1.01°$ (see Methods) and exhibits high-resistance band insulator (BI) states at full filling ($\nu_\mathrm{BG} = \pm 4$) of the moiré flat-bands, and a correlated insulator at half-filling on the electron-doped side ($\nu_\mathrm{BG} = 2$). BI states similarly emerge when the two FGs are swept independently at fixed $V_{\textrm{BG}}$ (Extended Data Fig.~\ref{figED2}), indicating that the small regions they control are uniformly twisted close to the magic angle. We observe superconducting (SC) domes at both electron and hole doping around half to full filling, as evidenced by a sudden drop in $R$ to near zero and the presence of a supercurrent (Extended Data Fig.~\ref{figED3}). The combination of highly insulating phases provided by BI gaps and superconducting phases enables the formation of gate-tunable quantum dot hybrid structures.

\subsection*{\label{sec3:superconducting_island} Superconducting island}

\begin{figure}
\includegraphics{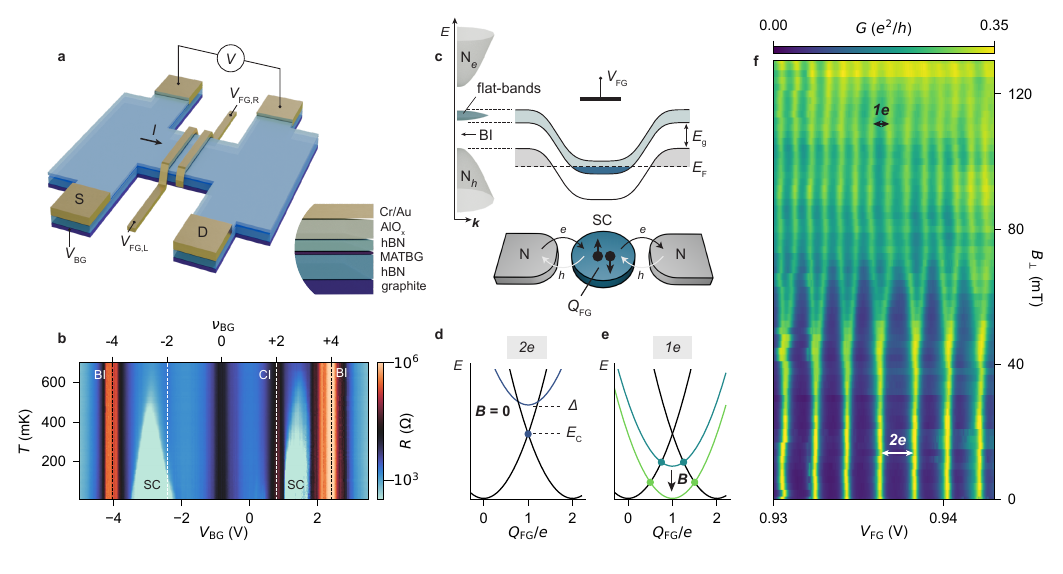}
\caption{\textbf{Device and superconducting island regime.} 
\textbf{a,} Device schematics and cross-section of the heterostructure (inset). MATBG is encapsulated in hBN, placed on a graphite back gate, and contacted by Cr/Au electrodes. The etched channel is traversed by two finger gates on top of AlO$_x$.
\textbf{b,} Four-probe bulk resistance $R$ as a function of back gate voltage $V_\mathrm{BG}$ and temperature $T$, showing superconducting (SC), band-insulating (BI), and a correlated-insulating (CI) state.
\textbf{c,} Top left: Schematic band structure of MATBG with the dispersive bands ($\textrm{N}_{e/h}$) separated from the flat-bands by the gap energy $E_g$ at the BI. Top right: Band edge profile along the etched channel. The island is formed as the finger gate locally tunes the MATBG from the dispersive hole band through the BI gap and into the superconducting flat-band regime. Bottom right: The superconducting (SC) island is coupled to normal (N) leads (bottom) and has total charge $Q_\textrm{FG}$.
\textbf{d,e,} Cooper-pair box energy diagrams. Odd-parity states are shifted by the gap $\Delta$. For $\Delta>E_\mathrm{C}$, odd states are inaccessible and transport is $2e$-periodic. A magnetic field $B$ reduces $\Delta$ and enables $1e$ transport.
\textbf{f,} Conductance $G$ versus $V_\mathrm{FG}$ and $B_\perp$ at $\nu_\mathrm{BG}=-4.4$, showing the evolution from $2e$-periodic Coulomb resonances at low fields through even-odd spacing (${50}~\textrm{mT} 	\lesssim B_\perp	\lesssim {70}~\textrm{mT}$) to $1e$-periodic oscillations when superconductivity is suppressed ($B_\perp \gtrsim~{80}~\textrm{mT}$).
} 
\label{fig1}
\end{figure}

We realize the superconducting island coupled to normal leads using the left FG (its voltage hereafter denoted as $V_{ \textrm{FG}}$), while keeping the right one at zero.
We position the electrochemical potential of the leads in the hole-doped dispersive band ($\nu_\textrm{BG} = -4.4$) and use $V_{ \textrm{FG}}$ to locally shift the state under the FG to a flat-band superconducting regime (Fig.~\ref{fig1}c, top). 
As the Fermi-level enters the BI gap, tunnel barriers develop at the FG--lead interfaces and thereby confine the superconducting island (Fig.~\ref{fig1}c, bottom).

In the Cooper-pair box model~\cite{Tuominen1992,elies1993}, the electrostatic energy of the island as a function of the gate-induced charge $Q_\textrm{FG} = C_\textrm{FG} V_\textrm{FG}$ (where $C_\textrm{FG}$ is the FG--Cooper-pair box capacitance) consists of a series of parabolas, each corresponding to a discrete number of charges $N_\textrm{FG}$ on the island (Fig.~\ref{fig1}d,e). 
Transport occurs at the degeneracy points where adjacent parabolas intersect, giving rise to conductance resonances. 
Odd-parity states require the presence of an unpaired quasiparticle and have their corresponding parabolas shifted up by the energy of the lowest quasiparticle excitation. 
For a superconductor without sub-gap states, this energy corresponds to the superconducting gap $\Delta$.
When $\Delta > E_\textrm{C}$, odd states are energetically inaccessible in low-bias linear transport, resulting in $2e$-periodic Coulomb oscillations (Fig.~\ref{fig1}d)~\cite{elies1993}. 
Quasiparticles contribute to transport once $\Delta$ is reduced below $E_\textrm{C}$, e.g. by applying a perpendicular magnetic field $B$ (Fig.~\ref{fig1}e).
This splits the $2e$-periodic resonances into an even-odd pattern, which evolves into uniformly spaced $1e$-periodic oscillations when superconductivity is fully quenched~\cite{Lu1996,Albrecht2016,Shen2018}.
In contrast, if low-energy quasiparticles states existed at zero field, as in the case of a nodal superconductor, odd states would participate in transport without suppressing superconductivity. This would preclude the even-odd parity effect and lead to $1e$-periodic transport already at zero magnetic field~\cite{mineev1994, Golubev1995}.

Figure~\ref{fig1}f shows that the measured conductance $G$ of the island at low source-drain bias $V_\textrm{SD} = {10} \mu\textrm{V}$ exhibits Coulomb oscillations as a function of $V_\textrm{FG}$. At zero magnetic field, the resonances are $2e$-periodic (with period $\delta V_\textrm{FG} \approx 2\textrm{ mV}$) and have vanishing conductance between peaks, consistent with a superconducting island in the $\Delta > E_\textrm{C}$ regime (Fig.~\ref{fig1}d). 
A perpendicular field $B_\perp \approx {50}~\textrm{mT}$ splits the $2e$ resonances into an alternating even-odd peak spacing as it lowers $\Delta$ below $E_\textrm{C}$. 
At higher fields ($B_\perp > {80}~\textrm{ mT}$), the oscillations become uniformly spaced and $1e$-periodic ($\delta V_\textrm{FG} \approx {1}~\textrm{mV}$), as expected from the suppression of superconductivity via an external magnetic field. 
We measure similar transitions of the period from $2e$ to $1e$ as a function of parallel magnetic field, temperature and bias voltage (Extended Data Fig.~\ref{figED4}), further confirming the superconducting origin of the period doubling. 
Additionally, we observe hints of oscillating even-odd spacing above the critical field of the island (Extended Data Fig.~\ref{figED5})~\cite{Albrecht2016,Shen2018,Valentini2022}. The same effects are independently reproduced using the right FG (Extended Data Fig.~\ref{figED6}).


The presence of nodes in the quasiparticle spectrum poses very stringent temperature requirements for the observation of the parity effect in nodal superconductors~\cite{Golubev1995}, not compatible with the extended range of temperatures at which the parity effect persists (see Extended Data Fig.~\ref{figED4}).
Therefore, the observation of robust $2e$-periodic transport is consistent with the presence of a superconducting island with a fully opened superconducting gap without accessible sub-gap quasiparticle states, as in $s$-wave islands and hard-gap hybrid systems~\cite{Albrecht2016,Shen2018,Pendharkar2021,Kanne2021}.
We attribute the presence of the well-defined gap to the small size of the island (${150}~\textrm{nm} \times {550}~\textrm{nm}$), which effectively confines transport to a local region of MATBG and reduces the impact of long-range twist-angle disorder~\cite{Uri2020}.


\subsection*{\label{sec4:N_to_S_dot} Tuning the island from normal to superconducting}

\begin{figure}
\centering
\includegraphics{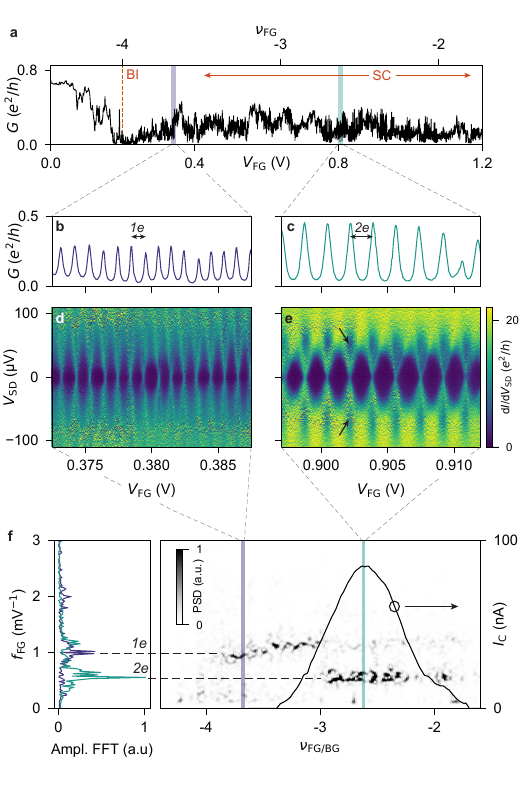}
\caption{\label{fig2} 
\textbf{Normal and superconducting Coulomb oscillations in the flat-bands.} 
\textbf{a,} $G$ as a function of $V_\textrm{FG}$ (bottom axis) and $\nu_\textrm{FG}$ (top axis), at fixed $\nu_\textrm{BG} = -4.4$. The orange lines mark the SC region and the BI state.
\textbf{b, c,} Zoom-ins of \textbf{a} showing $1e$-periodic Coulomb resonances in the normal regime (purple, \textbf{b}) and $2e$-periodic resonances in the superconducting regime (teal, \textbf{c}).
 \textbf{d, e,} Bias spectroscopy in the normal (\textbf{d}) and superconducting (\textbf{e}) regimes. Arrows in \textbf{e} indicate the regions of reduced conductance on resonance.
\textbf{f,} Fourier analysis of the Coulomb oscillations. Left: FFT amplitude of the traces in \textbf{b} (purple, $1e$) and \textbf{c} (teal, $2e$).
Right: power spectral density (PSD, grayscale) of the oscillations as a function of $\nu_\mathrm{FG}$, showing that the dominant $2e$ branch coincides with the superconducting region identified by the bulk critical current ($I_\textrm{c}$ versus $\nu_\mathrm{BG}$ overlaid onto the PSD map). 
}
\end{figure}

Having demonstrated a superconducting island, we exploit the tunability of MATBG to map the transition from a normal to a superconducting island as a function of $V_\textrm{FG}$. With the leads in the same normal state ($\nu_{\mathrm{BG}}=-4.4$), Fig.~\ref{fig2}a shows the conductance $G$ as a function of $V_\textrm{FG}$ (bottom axis), while the corresponding filling factor of the island ($\nu_\mathrm{FG}$) is indicated on the top axis. Here the FG locally tunes the Fermi-energy of the island from the dispersive hole band (at $V_\textrm{FG} = {0}~\textrm{V}$) across the BI state at $\nu_{\mathrm{FG}} = -4$ into the flat-bands. 
Upon entering the BI state (orange dashed line), the finite conductance of approximately $0.6e^2/h$ for $\nu_\textnormal{FG}<-4$ is strongly suppressed. With further increasing $V_\textrm{FG}$, whereby the Fermi level is shifted into the flat-bands, $G$ develops periodic Coulomb oscillations as evident from the zoomed-in traces in Figs.~\ref{fig2}b,c.

In the normal state (purple-shaded region in Fig.~\ref{fig2}a), the oscillations are spaced by $\delta V_\textrm{FG} \approx {1}~\textrm{mV}$ (Fig.~\ref{fig2}b). Having half the spacing of the resonances in the superconducting regime (Figs.~\ref{fig1}f and \ref{fig2}c) and showing no splitting under a magnetic field, they are consistent with single-carrier charging. From the size of the corresponding Coulomb diamonds in Fig.~\ref{fig2}d, we extract the single-electron charging energy $E_{\mathrm{C}} \approx {50}~\mu \textrm{V}$, compatible with the dimensions of the island. 

As $V_{\textrm{FG}}$ is tuned into the superconducting regime (SC range in Fig.~\ref{fig2}a), the system recovers the $2e$-periodic transport observed earlier (Fig.~\ref{fig2}c, compare with Fig.~\ref{fig1}f). The finite-bias differential conductance (Fig.~\ref{fig2}e) exhibits the characteristic features of a superconducting island~\cite{Tuominen1992,elies1993,Shen2018}.
On resonance, the differential conductance at finite bias is suppressed (black arrows) by bias-assisted quasiparticle transport beyond $|V_\textrm{SD}| = 2(\Delta- E_\textrm{C}) \approx {20}~\mu\textrm{V}$~\cite{elies1993, Hekking1993}. 
The $2e$-periodicity in $G$ persists up to $|V_\textrm{SD}| = 2\Delta \approx {100}~\mu \textrm{V}$ (see Extended Data Fig.~\ref{figED4}), beyond which above-gap quasiparticle states allow for $1e$ transport~\cite{elies1993}.
From these values, we extract $\Delta \approx {50}~\mu \textrm{V}$ and $E_\textrm{C} \approx {40}~\mu\textrm{V}$ and confirm that the island operates in the $\Delta \gtrsim E_\textrm{C}$ regime.

The Fourier spectra of the signals in Figs.~\ref{fig2}b,c (Fig.~\ref{fig2}f, left) show the distinct frequencies of Coulomb oscillations, $f_{1e}$ for the normal and $f_{2e} \approx f_{1e}/2$ for the superconducting state. 
Figure~\ref{fig2}f, right, shows the map of the power spectral density of the trace in Fig.~\ref{fig2}a as a function of $\nu_\mathrm{FG}$. 
At $\nu_\textrm{FG} \approx -3$, the dominant frequency abruptly halves, marking the transition into the superconducting state. 
Within the superconducting state (characterized by a finite bulk critical current $I_{\textrm{c}}$, trace in Fig.~\ref{fig2}f, right), $f_{2e}$ prevails as the main harmonic in the range $-3 \lesssim \nu_\textrm{FG} \lesssim -2.25$. This indicates the presence of the parity effect within this range, pointing to a fully gapped order parameter that persists throughout the superconducting dome.

\subsection*{\label{sec5:double_dot} Two superconducting islands}
\begin{figure}
\centering
\includegraphics{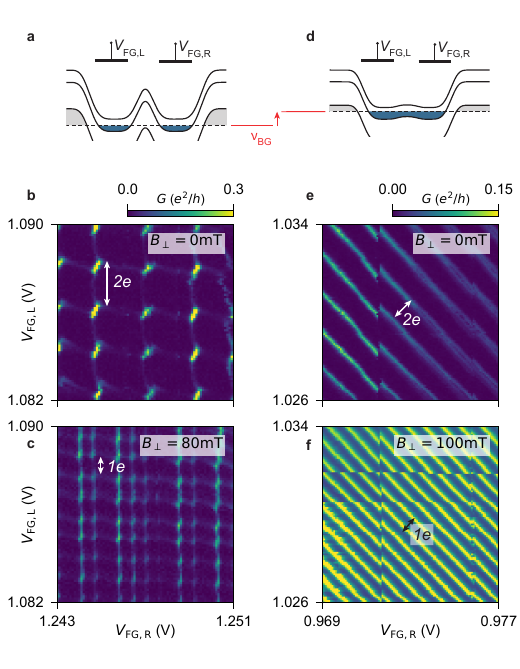}
\caption{\textbf{Double island regime.}
\textbf{a,b,} Band edge profiles for different values of $\nu_\textrm{BG}$. The finger gates (FGs) $V_\textrm{FG,L}$ and $V_\textrm{FG,R}$ define two independent islands (\textbf{a}) or a single island (\textbf{b}). 
\textbf{c,d,} Charge stability diagrams (CSDs) at $B_\perp = {0}~\textrm{T}$ in the double (\textbf{c}, $\nu_\mathrm{BG} = -4.7$) and single (\textbf{d}, $\nu_\mathrm{BG} = -4.5$) superconducting island regimes, showing $2e$-periodicity in both FGs.
\textbf{e,f,} CSDs at $B_\perp = {80}~\textrm{mT}$ (\textbf{e}) and $B_\perp = {100}~\textrm{mT}$ (\textbf{f}) for the same gate configurations with quenched superconductivity ($1e$).
\label{fig3}
}
\end{figure}

By using both finger gates ($V_\textrm{FG,R}$ and $V_\textrm{FG,L}$, see Fig.~\ref{fig1}a), we define a system with two superconducting islands in which each gate independently controls the charge occupancy of its respective island (Fig.~\ref{fig3}a). The region between the islands acts as a tunnel barrier and is primarily controlled by the BG. 

At $\nu_\text{BG} = -4.7$ (Fig.~\ref{fig3}b), the two islands are weakly coupled, and the charge stability diagram (CSD) exhibits the honeycomb pattern characteristic of a double-dot system~\cite{RevModPhys.75.1}. Conductance peaks occur only at $2e$-periodic triple points when the electrochemical potentials of even-parity states in both islands align.
At $B_\perp = {80}~\textrm{mT}$ (Fig.~\ref{fig3}c), the double-island transitions to $1e$-periodicity as the odd states become energetically available. While the resonances controlled by $V_\textrm{FG,L}$ are fully split, the alternating spacing and non-uniformity of the resonances controlled by $V_\textrm{FG,R}$ indicate a remaining parity effect in the right island. We then speculate that under this gate configuration, the zero-$B$ superconducting gap $\Delta$ is larger on the right island than on the left one.

As the BG is tuned to push the electrochemical potential toward the edge of the dispersive bands, the electrostatic profile becomes shallower, the region between the islands is pushed into the flat-bands, and the tunnel barrier disappears (Fig.~\ref{fig3}d). At $\nu_\textrm{BG} = -4.5$, the two islands merge into a single coherent superconducting island controlled equally by both finger gates. This is reflected by diagonal conductance resonances in the CSD (Fig.~\ref{fig3}e), which also exhibit the $2e$-to-$1e$ transition with magnetic field (Fig.~\ref{fig3}f, $B_\perp = {100}~\textrm{mT}$).

\subsection*{\label{sec6:double_dot}Proximitized quantum dot}

\begin{figure}
\includegraphics{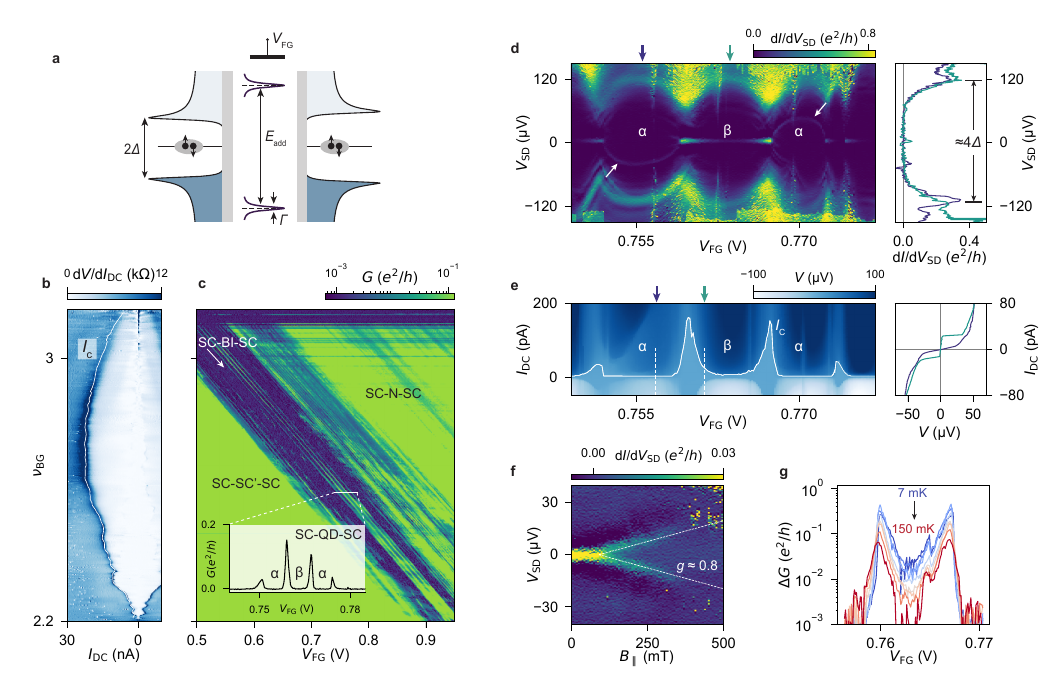}
\caption{
\textbf{Proximitized quantum dot.}
\textbf{a,} Schematic energy diagram of the quantum dot coupled to superconducting leads, with addition energy $E_\mathrm{add}$, tunnel coupling $\Gamma$, and superconducting gap $\Delta$.
\textbf{b,} Differential resistance $\textrm{d}V/\textrm{d}I_\mathrm{DC}$ of the bulk versus bias current $I_\textrm{DC}$ and $\nu_\textrm{BG}$. The finite critical current $I_\textrm{c}$ indicates superconductivity on the electron-doped side.
\textbf{c,} Conductance $G$ versus $V_\mathrm{FG}$ and $\nu_\mathrm{BG}$.
The labels SC-X-SC indicate the different states of the finger-gate region (X = SC', BI, QD, N) within the SC leads.
Inset: line cut of the four resonances with their charge parity labeled $\alpha$ and $\beta$. 
\textbf{d,} Left: Bias spectroscopy at fixed $\nu_\mathrm{BG}$. 
Right: Line cuts at the $\alpha$ (purple) and $\beta$ (teal) states. The conductance is suppressed within the superconducting gap ($\approx 4\Delta$ range).
\textbf{e,} Left: $V$-$I_\textrm{DC}$ characteristics as a function of $V_\mathrm{FG}$. There is a finite $I_\mathrm{c}$ (solid line) in the $\beta$ state that is suppressed in the $\alpha$ state.
Right panel: Line cuts at the same $V_\textrm{FG}$ as in (\textbf{e}).
\textbf{f,} Temperature dependence of the zero-bias excess conductance ($\Delta G = G - G_{\alpha}$, with $G_{\alpha}$ the average $G$ in $\alpha$) in the $\beta$.
\textbf{g,} In-plane magnetic field $B_\parallel$ dependence of the zero-bias conductance peak in $\beta$ state, which splits at $B_\parallel \sim {100}~\textrm{mT}$ with a slope corresponding to a $g$-factor $g \approx 0.8$.
}
\label{fig4}
\end{figure}

Exploiting the high tunability of MATBG, we access a complementary regime where the same device realizes a normal dot coupled to superconducting leads (SC–QD–SC, Fig.~\ref{fig4}a). 
To reach this regime, we first tune the leads into the superconducting state on the electron-doped side using the BG and confirm the presence of a finite bulk critical current $I_\textrm{c}$ when $V_{\textrm{FG,L/R} }$ = {0}~\textrm{V} (Fig.~\ref{fig4}b). 
Following an analogous tuning principle as in Fig.~\ref{fig1}c, we use only the left FG to bring the Fermi energy now from the flat-bands ($V_\textrm{FG}$ = {0}~\textrm{V}) into the electron-doped dispersive bands. 
The tunnel barriers that form when the Fermi level crosses the BI band gap define the SC–QD–SC system.

The conductance map as a function of $\nu_\textrm{BG}$ and $V_\textrm{FG}$ (Fig.~\ref{fig4}c) shows distinct transport regimes as the Fermi level in the FG region is tuned through the bands. When the Fermi level on the island remains in the flat-bands, the absence of a resistive barrier allows for the flow of supercurrent which results in the measured saturated conductance (green region labeled as SC-SC’-SC). Increasing $V_\textrm{FG}$ across the BI state suppresses both supercurrent and conductance (SC-BI-SC) and creates a gate-defined dot with clear conductance resonances that appear as diagonal lines.
Notably, we observe a set of four regularly spaced resonances grouped in a distinct bunch and spanning across the full SC range of the leads (SC-QD-SC, see inset for a line cut at fixed $\nu_\textrm{BG} = 2.6$). We speculate that its four-fold structure reflects the spin and valley degeneracy of MATBG and is consistent with the shell-filling patterns of carbon nanotubes~\cite{RevModPhys.87.703} and graphene quantum dots~\cite{eich2018}. A similar filling pattern is observed when the quantum dot is defined using the second (right) FG (Extended Data  Fig.~\ref{figED6}).

At a fixed filling of the leads $\nu_\textrm{BG} = 2.6$, the finite-bias differential conductance $\textrm{d}I/\textrm{d}V_{\textrm{SD}}$ of these four resonances exhibits a complex spectrum of sub-gap features (Fig.~\ref{fig4}d). These features are confined within a bias window $|V_\textrm{SD}| \lesssim 2\Delta$, as larger bias allows single-electron transport through quasiparticle states~\cite{DeFranceschi2010}. They display an even-odd pattern depending on the charge parity of the dot. We label the two parities $\alpha$ and $\beta$ without distinguishing even or odd, as transport measurements do not provide unambiguous evidence of the absolute electron number (Fig.~\ref{fig4}d and inset in c).
While the $\alpha$ states display curved sub-gap resonances that bend to zero energy at the resonance points and are not always electron-hole symmetric (white arrows in Fig.~\ref{fig4}d), these features do not appear in the $\beta$ states of opposite parity, which instead host a pronounced zero-bias conductance peak (ZBCP).

The transport properties and energy spectra of a hybrid quantum dot are governed by the interplay between the superconducting gap $\Delta$, the addition energy $E_\textrm{add}$, and the dot-lead tunnel coupling $\Gamma$~\cite{DeFranceschi2010}.
Tuning the state of the leads across the superconducting dome via $\nu_\textrm{BG}$ preserves the even-odd spectrum and rescales the bias window where the sub-gap features appear. This allows us to estimate $\Delta$ as a function of $\nu_\textrm{BG}$ (Extended Data Fig.~\ref{figED7}), the magnitude of which closely tracks the modulation of the critical current $I_\mathrm{c}$ across the superconducting dome, as expected for a gap feature.
For $\nu_\textrm{BG} = 2.6$ (Fig.~\ref{fig4}d), we extract $\Delta \approx {70}~\mu \textrm{eV}$, in good agreement with the BCS estimate of the critical temperature at this doping $T_\textrm{c} \approx \Delta/1.76 k_\textrm{B} \approx {460}~\textrm{mK}$ (compare with Fig.~\ref{fig1}b).

When the leads are driven to the normal state by an out-of-plane magnetic field exceeding the critical field, the differential conductance map reduces to conventional Coulomb diamonds (Extended Data Fig.~\ref{figED8}). From the extent of the diamonds, we extract an addition energy $E_\textrm{add} \approx {1}~\mu \textrm{eV}$, significantly larger than the estimated $\Delta$. We note that $E_\textrm{add}$ is also substantially larger than the $E_\textrm{C}$ extracted for the superconducting island, indicating a smaller effective dot size in the current regime. Furthermore, the observation of well-defined Coulomb diamonds indicates a weak tunnel coupling $\Gamma \ll E_{\textrm{add}}$. Together, this places the system in the regime $E_\mathrm{add} \gg \Delta, \Gamma$, with an undetermined ratio between $\Gamma$ and $\Delta$.

Current biased measurements in the same regime show the presence of a critical current that is enhanced when an energy level of the QD is aligned with the Fermi level of the superconducting leads (Fig.~\ref{fig4}e). 
As we established $E_\mathrm{add} \gg \Delta$, this supercurrent is presumably supported by high-order co-tunneling processes~\cite{DeFranceschi2010}. 
It exhibits an even-odd pattern as a finite $I_\textrm{c}$ remains in the $\beta$ states and we measure no supercurrent in the $\alpha$ states.
$I_\textrm{c}$ also decays in a highly asymmetric way when moving away from the resonances, vanishing sharply in $\alpha$ while smoothly decaying in $\beta$.

Assuming an odd/even electron occupancy in the $\alpha$/$\beta$ states, the curved sub-gap resonances in $\alpha$ (white arrow in Fig.~\ref{fig4}d) can be interpreted as Yu-Shiba-Rusinov excitations~\cite{Lee2014, doi:10.1126/science.abf1513}, which typically emerge when an unpaired spin is screened by the quasiparticles in the superconducting leads.
This parity assignment also explains the $I_\textrm{c}$ modulation in the context of a $\pi$-junction, where the spinful nature of the $\alpha$ state leads to a reduced and sign-reversed supercurrent~\cite{vanDam2006, Jorgensen2007}. 
This manifests in our measurements as a vanishing (presumably below the experimental resolution) $I_\textrm{c}$ in $\alpha$, and a sharp onset to finite $I_\textrm{c}$ at the charge transition to the $\beta$ state, corresponding to a switch in the direction of the supercurrent ($0$-$\pi$ junction).

The ZBCP has a bias voltage width $\delta V_\textrm{SD} \approx {15}~\mu\textrm{V}$ independent of $V_\textrm{FG}$ within $\beta$ (Extended Data Fig.~\ref{figED9}). This width is broader than $k_\textrm{B} T \lesssim {1}~\mu \textrm{eV}$, and also exceeds the finite voltage drop due to phase diffusion~\cite{Jorgensen2007} ($\delta V \lesssim {5}~\mu \textrm{V}$ and also $V_\textrm{FG}$-dependent, Extended Data Fig.~\ref{figED9}).
Furthermore, an in-plane magnetic field $B_\parallel$ linearly splits the ZBCP above $B_\parallel \approx {100}~\textrm{mT}$ according to $e|V_\textnormal{SD}|=g~\mu_\textnormal{B}B_\parallel$ with a $g$-factor $g \approx 0.8$ (Fig.~\ref{fig4}f). This splitting, together with the observed bias width, points towards a spinful state pinned at zero energy.

The temperature evolution of the ZBCP (Fig.~\ref{fig4}g) shows that increasing the temperature reduces the zero-bias conductance and that the ZBCP vanishes at temperatures above $T \approx 120~\textrm{mK}$, significantly below the critical temperature of the superconductor at this doping ($T_\textrm{c} \approx {400}~\textrm{mK}$). 
This trend, similar to that expected for a Kondo resonance~\cite{Goldhaber-Gordon1998}, also excludes thermally activated quasiparticles as the origin of the ZBCP.

\subsection*{\label{sec7:conclusion_and_outlook} Discussion and outlook}
We demonstrate that mesoscopic transport through a Coulomb-blockaded superconducting island provides a sensitive probe of the low-energy quasiparticle spectrum in MATBG, bridging the gap between local\cite{Oh2021,Kim2022,Kim2026} and spatially averaged~\cite{Park2026} tunneling spectroscopy measurements.
The observed $2e$-periodic transport, together with its magnetic field, bias spectroscopy and temperature dependence, is incompatible with zeros in the order parameter and consistent with a fully gapped superconducting order parameter.

The same architecture provides a foundation for hybrid superconducting-semiconducting devices in moiré materials. Their tunability between normal, insulating, and superconducting phases allows superconducting islands and proximitized quantum dots to be realized within a single crystal. This monolithic and reconfigurable platform can be scaled to more advanced hybrid architectures like hybrid dot arrays. 
In future experiments, independent control of each lead between superconducting and normal states will enable versatile engineering of the proximitized-dot spectrum and provide new opportunities for spectroscopy of the superconducting gap and quasiparticle excitation spectrum.






\clearpage 

%
\bibliography{mybibliography} 
\bibliographystyle{sciencemag}

%
%
%
%
%
%


\section*{Acknowledgments}
The authors thank Eran Sela, Yigal Meir, Gianni Blatter, Fabrizio Nichele and Christian Sch{\"o}nenberger for stimulating discussions and comments. We thank Peter M\"{a}rki, Thomas B\"{a}hler, and the staff of the ETH cleanroom facility FIRST for technical support. 
\paragraph*{Funding:}
We acknowledge financial support by the European Graphene Flagship Core3 Project, H2020 European Research Council (ERC) Synergy Grant under Grant Agreement 951541.
K.W. and T.T. acknowledge support from the CREST (JPMJCR24A5), JST and World Premier International Research Center Initiative (WPI), MEXT, Japan.
\paragraph*{Author contributions:}
A.M., A.D., T.I. and K.E. conceived the experiment. 
A.M. fabricated the device and performed the measurements.
A.M. and A.O. interpreted the data with inputs from M.P, C.G.A, T.I. and K.E.
T.T. and K.W. supplied the hBN crystals.
A.M. wrote the manuscript with inputs from A.D. and comments from all authors.

\paragraph*{Competing interests:}
The authors declare no competing interests.
\paragraph*{Data, code and materials availability:}
The data that support the findings of this study will be made available online through the ETH Research Collection.


\subsection*{Supplementary materials}
Materials and Methods\\
Figs. S1 to S9\\


\newpage


\renewcommand{\thefigure}{S\arabic{figure}}
\renewcommand{\thetable}{S\arabic{table}}
\renewcommand{\theequation}{S\arabic{equation}}
\renewcommand{\thepage}{S\arabic{page}}
\setcounter{figure}{0}
\setcounter{table}{0}
\setcounter{equation}{0}
\setcounter{page}{1} 


\begin{center}
\section*{Supplementary Materials for\\ \scititle}
Alexandra~Mestre-Tor\`a$^{1,2\ast}$,
Marta~Perego$^{1,2}$,
Clara~Galante~Agero$^{1,2}$,
Takashi~Taniguchi$^{3}$,\and
Kenji~Watanabe$^{4}$,
Thomas~Ihn$^{1,2}$,
Klaus~Ensslin$^{1,2}$
Artem~O.~Denisov$^{1,2}$\and
\\
\small$^{1}$Laboratory for Solid State Physics, ETH Z\"urich, Z\"urich, Switzerland.\\
\small$^{2}$Quantum Center, ETH Z\"urich, Z\"urich, Switzerland.\\
\small$^{3}$Research Center for Materials Nanoarchitectonics, National Institute for Materials Science, Tsukuba, Japan.\\
\small$^{4}$Research Center for Electronic and Optical Materials, National Institute for Materials Science, Tsukuba, Japan.\\
\small$^\ast$Corresponding author. Email: amestre@phys.ethz.ch
\end{center}

\subsubsection*{This PDF file includes:}
Materials and Methods\\
Figures S1 to S9\\

\newpage


\subsection*{Materials and Methods}

\subsubsection*{Device fabrication}
The 2D-heterostructure is assembled using a dry-transfer process with a polydimethylsiloxane/polycarbonate stamp and consists, from top to bottom, of a thin (thickness {5}\textrm{nm}) top hBN crystal, MATBG, a bottom hBN ({5}\textrm{nm}) and a graphite flake. The initial monolayer graphene is cut with a tungsten needle with a tip diameter of ${2}\mu{m}$ controlled by a micromanipulator.
Ohmic contacts and metal gates are fabricated with a lift-off process by electron beam lithography and metal deposition (Cr/Au). We make edge contacts by using reactive ion etching (RIE) with CHF$_3$ to etch hBN and graphene before metal deposition. We also use RIE to define the mesa structure. The top gate layer is insulated from the graphene edges of the etched structure by a ${15}\textrm{nm}$ layer of AlO$_{x}$ deposited by atomic layer deposition at ${150}\textrm{C}$. Optical and AFM images of the device are presented in Extended Data Fig.~\ref{figED1}.

\subsubsection*{Measurement setup}

\begin{figure}
\includegraphics{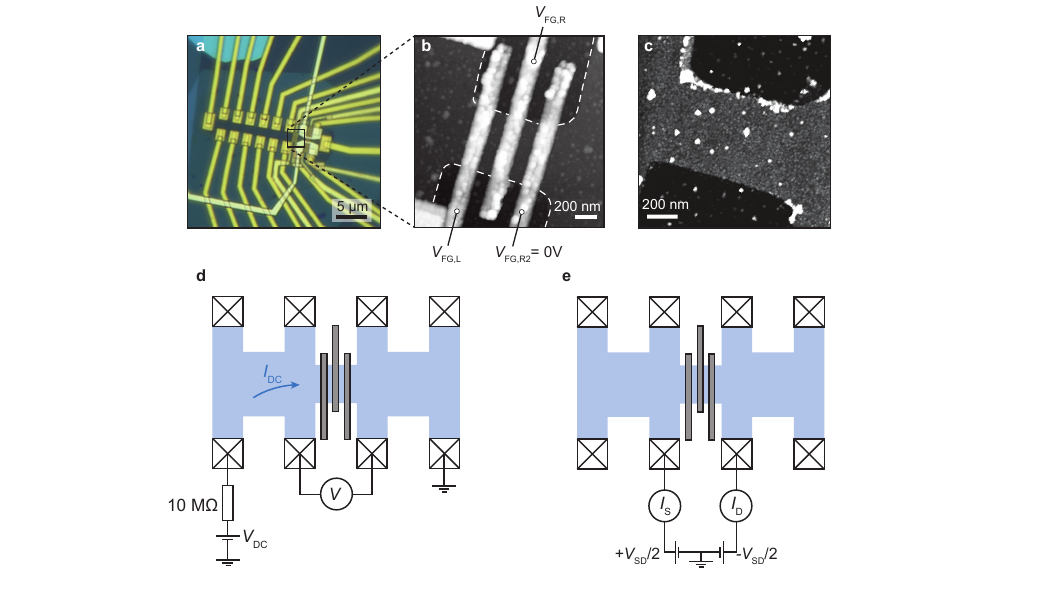}
\caption{\textbf{Sample and measurement setups.} 
\textbf{a,} Optical image of the sample. 
\textbf{b,} AFM image of the device. The white dashed line highlights the edge of the etched channel. The finger gate layer consists of three finger gates, of which we only use the two left ones ($V_\mathrm{FG,L}$ and $V_\mathrm{FG,R}$). The third finger gate ($V_\mathrm{FG,R2}$) is kept at ${0}~\textrm{V}$ in all measurements.
\textbf{c,} AFM image of the etched channel before the aluminum oxide layer.
\textbf{d,} Schematic of the four-terminal current bias measurements. The constant current $I_\textrm{DC}$ is sourced by applying a constant bias $V_\textrm{DC}$ through a $10~\textrm{M}\Omega$ series resistor.
\textbf{e,} Schematic of the voltage-biased measurements. The sample is symmetrically biased with a source-drain voltage bias $V_\textrm{SD}$. The current flowing through the structure is measured both at the source and drain contacts ($I_\textrm{S/D}$) with an IV converter (not illustrated in the schematic).
\label{figED1}
}
\end{figure}

\begin{figure}
\includegraphics{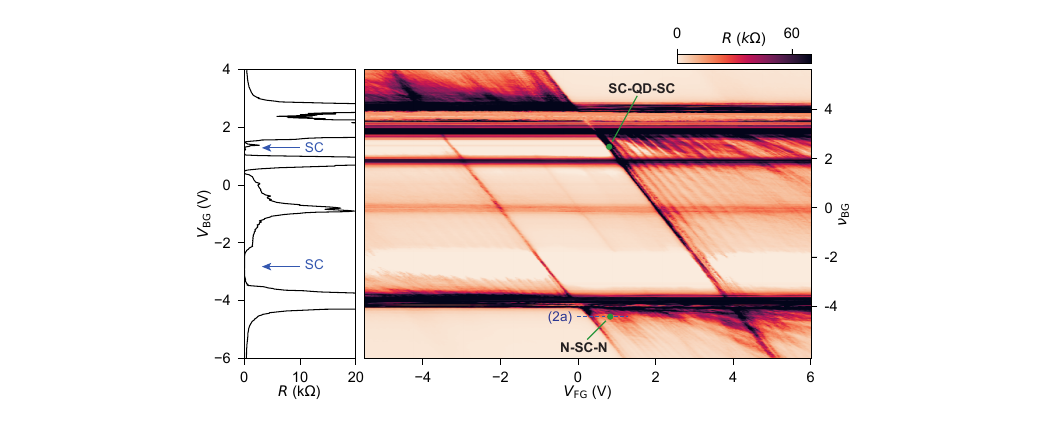}
\caption{\textbf{Gate-gate resistance map.} 
Left panel: Bulk four-terminal resistance $R$ as a function of $V_\textrm{BG}$ at $V_\textrm{FG}={0}~\textrm{V}$. Regions of vanishing resistance are indicated with the label SC and correspond to the superconducting phases on the electron- and hole-side. 
Right panel: Four-terminal resistance $R$ as a function of $V_\textrm{FG}$ and $V_\textrm{BG}$, plotted alongside with the corresponding bulk filling factor $\nu_\textrm{BG}$. Horizontal features correspond to resistive states in the bulk at integer filling factors of the moiré unit cell. Diagonal features originate from the small region controlled by both the finger gate and the back gate. Green dots indicate the points where the superconducting island (N-SC-N) and the proximitized quantum dot (SC-QD-SC) regimes are realized. The blue dashed line marks the voltage range of the trace in Fig.~\ref{fig2}a. 
\label{figED2}
}
\end{figure}

\begin{figure}
\includegraphics{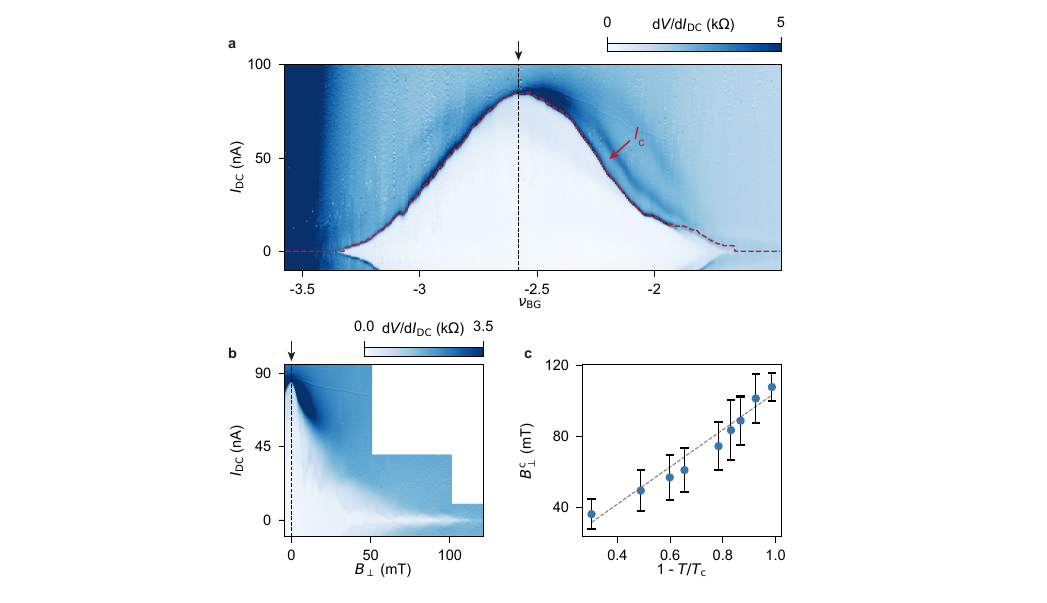}
\caption{\textbf{Characterization of bulk superconductivity.} \textbf{a,} Differential resistance $\mathrm{d}V/\mathrm{d}I_\mathrm{DC}$ as a function of bias current $I_\textrm{DC}$ and bulk filling factor $\nu_\mathrm{BG}$ in the superconducting regime on the electron-doped side. 
The red dashed line traces the critical current $I_\mathrm{c}$ (plotted in Fig.~\ref{fig2}f).
\textbf{b,} $\mathrm{d}V/\mathrm{d}I_\mathrm{DC}$ as a function of perpendicular magnetic field $B_\perp$ at optimal doping ($\nu_\textrm{BG} = -2.6$, black vertical dashed line in \textbf{a}). The critical field $B_\perp^\textrm{C}$ at this doping is  $B_\perp^\textrm{C} \approx {110}~\textrm{mT}$.
\textbf{c,} $B_\perp^\textrm{C}$ as a function of the reduced temperature $(1-T/T_\textrm{c})$, where $T_\textrm{c}$ is the superconducting critical temperature at the optimal filling factor. $B_\perp^\textrm{C}$ is extracted at the $50\%$ of the normal-state resistance with error bars representing the uncertainty of the superconducting transition determined by the $40\%$ and $60\%$ of the normal-state value. The dashed gray line is a fit of the expression $B_\perp^\mathrm{C} = \Phi_0/(2 \pi \xi_0^2) (1 - T/T_\mathrm{c})$, where $\Phi_0$ is the magnetic flux quantum and $\xi_0$ is the Ginzburg–Landau coherence length at zero temperature. From the fitting, we extract $T_\textrm{c} = {579}~\textrm{mK}$ and $\xi_0 = {57}~\textrm{nm}$.
}
\label{figED3}
\end{figure}

\begin{figure}
\includegraphics{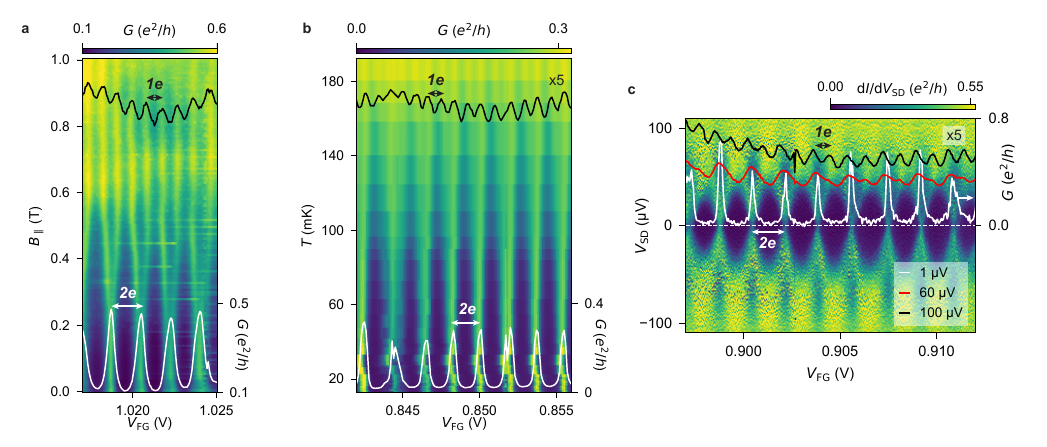}
\caption{\textbf{$2e$-to-$1e$ transition in the superconducting island regime with in-plane magnetic field, temperature and bias voltage.}
\textbf{a,} Conductance $G$ of the superconducting island as a function of $V_\textrm{FG}$ and in-plane magnetic field $B_\parallel$. The traces are a linecuts at $B_\parallel = {0}~\textrm{T}$ (white, showing $2e$-periodic transport) and $B_\parallel = {0.8}~\textrm{T}$ (black, showing $1e$-periodic transport).
\textbf{b,} $G$ as a function $V_\textrm{FG}$ and temperature $T$. The traces are a linecuts at $T = {20}~\textrm{mK}$ (white, $2e$-periodic) and $T = {180}~\textrm{mK}$ (black, $1e$-periodic and scaled by a factor $\times 5$).
The observation of the parity effect at high temperatures ($>150~\textrm{mK}$) is not compatible with presence of nodes in the quasiparticle spectrum. Theory predicts that parity-protected transport only survives up to $T_\textrm{cr} \sim \Delta / \sqrt{N_S \ln{100 N_S}}$ (for $d$-wave superconductor), where $N_S$ is the number of superconducting electrons~\cite{Golubev1995}. Using $\Delta \approx 50~\mu\textrm{eV}$ and $N_S \sim 2000$ (estimated from the area of the device and the superconducting carrier density), we estimate $T_\textrm{cr} \lesssim 5~\textrm{mK}$, more than an order of magnitude below the temperatures over which we observe the parity effect.
\textbf{c,} $G$ as a function $V_\textrm{FG}$ at bias voltages $V_\mathrm{SD} = {10}{}, {60}{}, {100}~\mu \textrm{V}$, overlaid on the corresponding differential conductance $\textrm{d}I/\textrm{d}V_\textrm{SD}$ versus $V_\textrm{SD}$ map. $G$ is $2e$-periodic at low bias ($V_\mathrm{SD} = {10}~\mu \textrm{V}$) and the $1e$-periodicity emerges for $V_\mathrm{SD} \ge {100}~\mu \textrm{V}$, coinciding with the onset of quasiparticle transport when $V_\mathrm{SD} \ge 2\Delta$. Traces are vertically offset for visibility. 
\label{figED4}
}
\end{figure}

\begin{figure}
\includegraphics{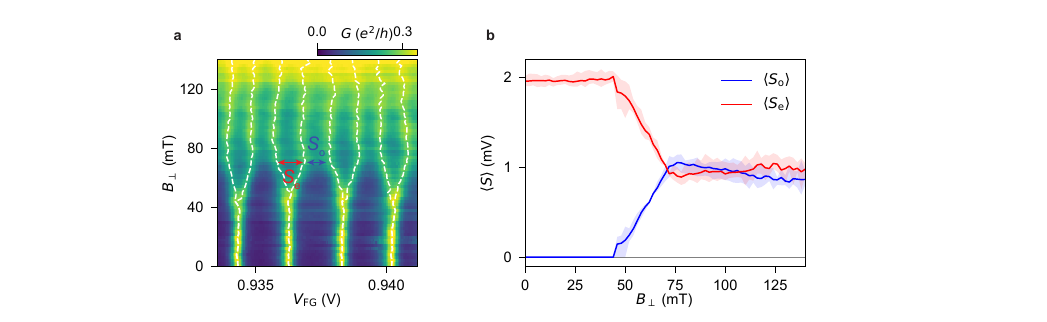}
\caption{\textbf{Even-odd modulation of the peak spacing.} 
\textbf{a, } Conductance $G$ as a function of $V_\mathrm{FG}$ and out-of-plane magnetic field $B_\perp$, with the white dashed line tracing the extracted peak positions. The red and blue arrows indicate, respectively, the even $S_\textrm{e}$ and odd $S_\mathrm{o}$ spacings. 
\textbf{b, } Average even/odd $\langle S_\mathrm{e} \rangle$/$\langle S_\mathrm{o} \rangle$ spacing as a function of $B_\perp$, showing a modulation above the $2e$ to $1e$ transition.}
\label{figED5}
\end{figure}

Measurements are performed in a dry $^3$He/$^4$He dilution cryostat with a base temperature of 10 mK. 
For the finite-bias (high impedance) measurements, we apply a symmetric DC source-drain bias $V_\textrm{SD}$ across the channel and probe the resulting current $I$ with a built-in-house current-to-voltage converter with a feedback resistor of ${1}\textrm{G}\Omega$ and a Hewlett Packard 34401A multimeter.
The conductance $G$ is obtained after subtracting the series resistance of ${25}\textrm{G}\Omega$ from the measurement lines. Differential conductance is computed via numerical differentiation of the $I(V_\textrm{SD})$ traces.
For low-impedance measurements, we use a four-terminal current bias configuration. We source a DC current $I$ by applying a voltage from a digital-to-analog converter through a ${10}\textrm{G}\Omega$ series resistance. The resulting voltage drop $V$ is amplified with an in-house-built voltage amplifier and sensed with a Hewlett-Packard 34401A multimeter. Schematics of the two measurement setups are presented in Extended Data Fig.~\ref{figED1}.

\subsubsection*{Moir\'e filling factor and twist angle extraction}
We determine the filling factor of the Moir\'e superlattice of the bulk ($\nu_\mathrm{BG}$) from the gate voltage $V_\mathrm{BG}$.
First, we extract the carrier density at full filling of the Moir\'e unit cell via a parallel-plate capacitor model: 
$n_\textrm{BI} = C_\textrm{BG} (V_\textrm{BG}^\textrm{BI} - V_\textrm{BG}^\textrm{CN})/e$,
where $C_\textrm{BG}$ is the back-gate capacitance (estimated from the thickness and permittivity of the hBN dielectric layer), $V_\textrm{BG}^\textrm{BI}$ and $V_\textrm{BG}^\textrm{CN}$ are the BG voltages at the band insulator (BI) and the charge neutrality (CN) points and $e$ is the elementary charge.
By normalizing the gate-induced charge to $n_\mathrm{BI}$, we extract the filling factor as:
$\nu_\textrm{BG} = \frac{4}{n_\textrm{BI}}  C_\textrm{BG} (V_\textrm{BG} - V_\textrm{BG}^\textrm{CN})$.

From $n_\textrm{BI}$ we also estimate the average twist angle of the sample as $\theta = \sqrt{\frac{ \sqrt{3} }{8} n_\textrm{BI} a^2}$, where $a$ is the lattice constant of graphene.



\begin{figure}
\includegraphics{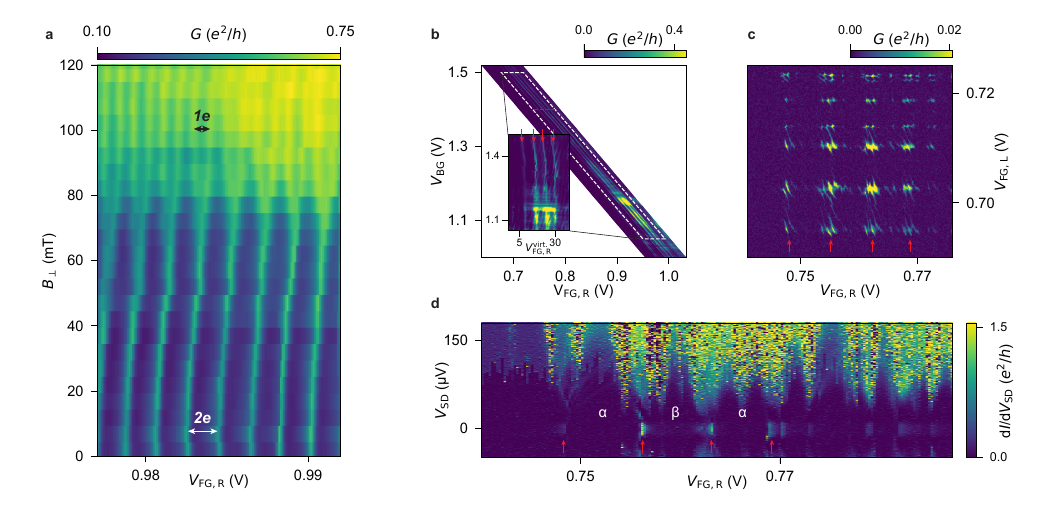}
\caption{\textbf{Second gate reproducibility.}
\textbf{a,} Superconducting island. Conductance $G$ as a function of perpendicular field $B_\perp$ of the superconducting island defined using the right (second) finger gate $V_\textrm{FG,R}$. Consistent with the observations on the left finger gate ($V_\textrm{FG,L}$), the system exhibits $2e$-periodic resonances at zero field that split into $1e$-periodic oscillations at $B_\perp \gtrsim {80}~\textrm{mT}$. 
\textbf{b,c,d} Proximitized quantum dot. \textbf{b,} $G$ as a function of $V_\textrm{BG}$ and $V_\textrm{FG,R}$ in the proximitized dot regime showing a charge stability map analogous to Fig.~\ref{fig4}c, also exhibiting four distinct resonance lines (indicated with a red arrow) that span through the charge map.
The dashed white box outlines the region shown in the inset, where the same data is plotted against a virtual gate voltage $V_{\text{FG,R}}^{\text{virt.}}$ (in units of mV) for visualization purposes. 
\textbf{c,} $G$ as a function of $V_\textrm{FG,L}$ and $V_\textrm{FG,R}$ at a fixed $V_\textrm{BG} = {1.4}~\textrm{V}$ (red dashed line in \textbf{b}, which corresponds to $\nu_\textrm{BG} = 2.8$) demonstrating that each set of four resonances are independently controlled by each gate. The red arrows indicate the four resonances in \textbf{b}.
\textbf{d,} Differential conductance $\mathrm{d}I/\mathrm{d}V_\mathrm{SD}$ as a function of source-drain bias $V_\mathrm{SD}$ and $V_\mathrm{FG,R}$ at a fixed $V_\textrm{BG} = {1.4}~\textrm{V}$ showing a finite conductance at zero bias in the $\beta$ states and a complete suppression of $G$ in the $\alpha$ states.
}
\label{figED6}
\end{figure}

\begin{figure}
\includegraphics{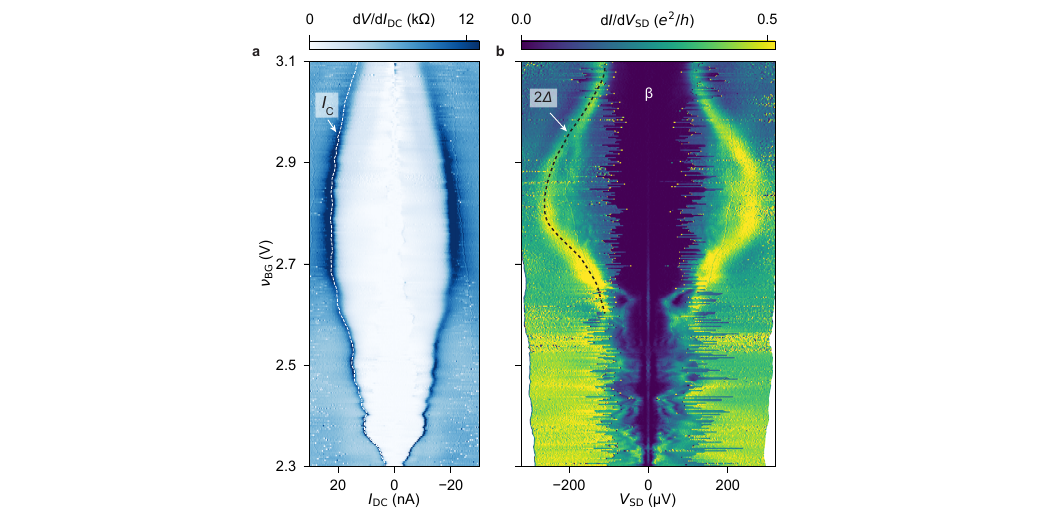}
\caption{\label{figED6} \textbf{Superconducting gap and bulk critical current.} 
\textbf{a,} Differential resistance $\mathrm{d}V/\mathrm{d}I_\mathrm{DC}$ as a function of bias current $I_\textrm{DC}$ and bulk filling factor $\nu_\mathrm{BG}$ in the superconducting regime on the electron-doped side. The white dashed line indicates the critical current $I_\mathrm{c}$.
\textbf{b,} Differential conductance $\mathrm{d}I/\mathrm{d}V_{\mathrm{SD}}$ as a function of $\nu_\mathrm{BG}$ in the same range as in \textbf{a} with $V_\textrm{FG,L}$ realizing a SC-QD-SC in the $\beta$ state. The map captures the dependence of the petrol trace in Fig. 4d as a function of $\nu_\mathrm{BG}$. The modulation of the superconducting gap with $\nu_\mathrm{BG}$ is highlighted with the black dashed line and closely follows the evolution of $I_\textrm{c}$ for $\nu_\mathrm{BG} \gtrsim 2.6$.
}
\label{figED7}
\end{figure}

\begin{figure}
\includegraphics{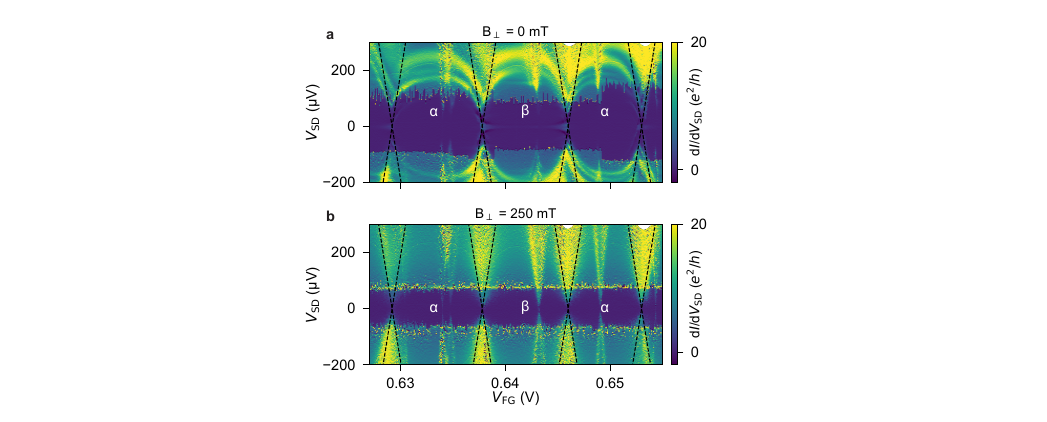}
\caption{\textbf{SC-QD-SC regime for a second $\nu_\mathrm{BG}$.}
\textbf{a,} Differential conductance $\mathrm{d}I/\mathrm{d}V_{\mathrm{SD}}$ as a function of $V_\mathrm{SD}$ and $V_\mathrm{FG}$ in the SC-QD-SC regime with the leads at a filling factor $\nu_\mathrm{BG} = 2.9$. For this filling factor, the sub-gap features are confined within a bias window $\pm 2\Delta \approx 250~\mu \textrm{eV}$. The $\beta$ state also shows a zero-bias conductance peak that is absent in the $\alpha$ state.
\textbf{b,} The same map at a perpendicular magnetic field $B_\perp = {250}~\textrm{mT}$, above the critical field. 
The sub-gap and zero-bias features vanish as the superconducting gap is suppressed and we observe standard Coulomb diamonds. 
The black dashed lines, serving as a guide to the eye, highlight the Coulomb diamond boundaries and from which we extract a finger gate lever arm $\alpha_\textrm{FG} \approx 0.12$ and an addition energy $E_\mathrm{add}\approx {1}~{meV}$. 
We speculate that the extra diamond-like shaped features arise from a small spurious dot in parallel with the studied proximitized dot. 
}
\label{figED8}
\end{figure}

\begin{figure}
\includegraphics{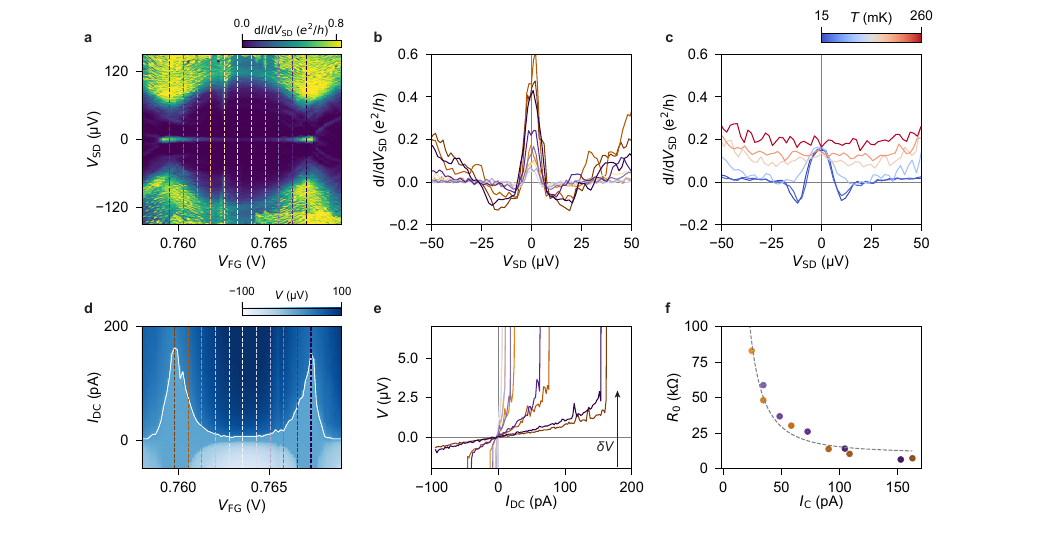}
\caption{\textbf{Finger gate dependence of the zero-bias feature.}
 \textbf{a,} Differential conductance $\mathrm{d}I/\mathrm{d}V_{\mathrm{SD}}$ in the $\beta$ diamond.
 \textbf{b,} Line cut at the $V_{\mathrm{FG}}$ values indicated with dashed lines in \textbf{a}. The traces show a conductance peak at zero bias followed by a region of negative differential conductance at around $\pm {20}~\mu \textrm{V}$. The width of the peak in $V_{\mathrm{SD}}$ ($\delta V_{\mathrm{SD}} \approx {15}~\mu \textrm{V}$) does not significantly change within the $\beta$ state. 
 \textbf{c,} Temperature dependence of the zero-bias conductance peak, showing that the feature vanishes at $T \approx {50}~\textrm{mK}$.
 \textbf{d,} $V-I_{\mathrm{DC}}$ characteristics in the $\beta$ state. The white line outlines the critical current.
 \textbf{e,} Line cuts of the $V-I_{\mathrm{DC}}$ map for the same $V_{\mathrm{FG}}$ values as in \textbf{a}. The finite slope around $I_{\mathrm{DC}} \approx 0$ is $V_{\mathrm{FG}}$-dependent, and we attribute it to phase diffusion arising from the high-resistance junction. The width in voltage drop corresponding to this phase diffusion branch is $\delta V \lesssim {6}~\mu \textrm{V}$.
 \textbf{f,} Scatter plot of the resistance of the phase diffusion branch $R_0$ as a function of the critical current $I_\textrm{c}$. The resistance values are extracted from a linear fit to the $V-I_{\mathrm{DC}}$ curves below $I_\textrm{c}$. In the limit where the Josephson energy $E_\mathrm{J} \ll k_\mathrm{B} T$, the expression $R_0 = 2 Z_1 (k_\mathrm{B} T / E_\textrm{J})^2$ relates $I_\textrm{c}$ and $R_0$. This formula fits the experimental data (gray dashed line) and, assuming an electron temperature of $T = {10}~\textrm{mK}$, it yields an effective environmental impedance of $Z_1 = {133}~\Omega$.
 }
 \label{figED9}
\end{figure}

\end{document}